\documentclass[aps,pra,amsmath,amssymb,twocolumn,superscriptaddress,showpacs,floatfix]{revtex4-1}
\usepackage{ifpdf}
 \newif\ifpdf
\ifx\pdfoutput\undefined
   \pdffalse
\else
   \pdfoutput=1
   \pdftrue
\fi
\ifpdf
   \usepackage{graphicx}
   \usepackage{epstopdf}
   \usepackage{color}
   \usepackage{verbatim}
\else
   \usepackage{graphicx}
\fi
\usepackage{bm}
\usepackage{srctex}
\usepackage{hyperref}

\DeclareMathOperator{\EFMpFM}  {\mathit{E}^{p-spin \,\mathrm{\scriptscriptstyle{FM}}}_{\mathrm{\scriptscriptstyle{FM}}}}
\DeclareMathOperator{\EAFMpFM}{\mathit{E}^{p-spin \,\mathrm{\scriptscriptstyle{FM}}}_{\mathrm{\scriptscriptstyle{AFM}}}}
\DeclareMathOperator{\ENEMpFM}{\mathit{E}^{p-spin \,\mathrm{\scriptscriptstyle{FM}}}_{\mathrm{\scriptscriptstyle{NEM}}}}

\DeclareMathOperator{\EFMpAFM}  {\mathit{E}^{p-spin \,\mathrm{\scriptscriptstyle{AFM}}}_{\mathrm{\scriptscriptstyle{FM}}}}
\DeclareMathOperator{\EAFMpAFM}{\mathit{E}^{p-spin \,\mathrm{\scriptscriptstyle{AFM}}}_{\mathrm{\scriptscriptstyle{AFM}}}}
\DeclareMathOperator{\ENEMpAFM}{\mathit{E}^{p-spin \,\mathrm{\scriptscriptstyle{AFM}}}_{\mathrm{\scriptscriptstyle{NEM}}}}

\begin{document}


\title{Quantum phase transitions and the degree of nonidentity in the system with two different species of vector bosons}

\author{A.~M.~Belemuk}
\affiliation{Institute for High Pressure Physics, Russian Academy of Sciences, Moscow (Troitsk) 108840, Russia}
\affiliation{Department of Theoretical Physics, Moscow Institute of Physics and Technology  (State University), Moscow 141700, Russia}

\author{N.~M.~Chtchelkatchev}
\affiliation{Institute for High Pressure Physics, Russian Academy of Sciences, Moscow (Troitsk) 108840, Russia}
\affiliation{Department of Theoretical Physics, Moscow Institute of Physics and Technology  (State University), Moscow 141700, Russia}
\affiliation{L.D. Landau Institute for Theoretical Physics, Russian Academy of Sciences, Moscow  119334, Russia}
\affiliation{Institute of Metallurgy, Ural Branch, Russian Academy of Sciences, Ekaterinburg 620016, Russia}

\author{A.~V.~Mikheyenkov}
\affiliation{Institute for High Pressure Physics, Russian Academy of Sciences, Moscow (Troitsk) 108840, Russia}
\affiliation{Department of Theoretical Physics, Moscow Institute of Physics and Technology (State University), Moscow 141700, Russia}
\affiliation{National Research Center ``Kurchatov Institute'', Moscow 123182, Russia}

\author{K.I. Kugel}
\affiliation{Institute for Theoretical and Applied Electrodynamics, Russian Academy of Sciences, Moscow 125412, Russia}
\affiliation{National Research University Higher School of Economics, Moscow 101000, Russia}

\date{\today}

\begin{abstract}
We address the system  with two species of vector bosons in an optical lattice.  In addition to the the standard parameters characterizing such a system, we are dealing here with the ``degree of atomic nonidentity'', manifesting itself in the difference of tunneling amplitudes and on-site Coulomb interactions. We  obtain a cascade of quantum phase transitions occurring with the increase in the degree of atomic nonidentity. In particular, we show  that the phase diagram for strongly distinct atoms is qualitatively different from  that for (nearly) identical atoms considered earlier. The resulting phase diagrams evolve from the images similar to the  ``J.~Mir\'o-like paintings'' to ``K.~Malewicz-like'' ones.
\end{abstract}


\maketitle


\section{Introduction}\label{sec.intro}
Experimental research of ultracold atoms in optical lattices have dramatically expanded the possibilities of a tunable simulation of quantum many-body physics~\cite{Greiner2002Nature,Kohl2005PRL,kawaguchi2012spinor,hauke2012,
baier2016Science,Bermudez2017PRB,mazurenko2017Nature}. Moreover,  ultracold atoms open the path to the parameter range that is hardly possible or even
impossible to achieve in the natural condensed matter systems~\cite{Meacher1998,Bruder1998PRL,Ovchinnikov2000,Corboz2013PRX}.

The typical example is the system of vector bosons. This case corresponds to Bose--Hubbard model that is absent in the standard solid state theory~\cite{krutitsky_ultracold_2016}. The situation becomes even more intriguing, when the problem implies some additional nontrivial parameters. In our case, we have multiple vector boson species~\cite{belemuk2017PRB}. Then,  in addition to the standard parameters, there appear nontrivial ones related to the   ``degree of atomic nonidentity'': the difference of tunneling amplitudes and on-site interactions.

Vector two-species bosons in optical lattices  are characterized by the following parameters: hopping amplitudes $t_{\alpha}$, where $\alpha=1,2$ labels different bosons, $U_{\alpha,\alpha'}$  --- on-site interactions and spin-channel interaction parameters $U_s$~\cite{krutitsky_ultracold_2016,belemuk2017PRB}. In recent paper, we have considered the simplest limiting case of nearly identical bosons~\cite{belemuk2017PRB} in the Mott insulating state: $U_{12}\simeq U_{11}\simeq U_{22}=U_0$ and $t_1\simeq t_2\ll U_0$. This model differs from the case of perfectly identical bosons by the absence of cross-tunneling term: tunneling with the change of boson identity was forbidden. It has been shown in~\cite{belemuk2017PRB}  that the model can be reduced to the Kugel--Khomskii~\cite{Kugel1982UFN} type spin--pseudospin model (where pseudospin labels different bosons). The assumption about perfectly identical bosons have lead to rather simple and intuitively expected phase diagram with one quantum phase transition near $U_s=0$~\cite{belemuk2017PRB}.

Here, we investigate  two species of cold vector bosons (ultacold bosonic atoms) in an optical lattice in the Mott insulating state ($U_{\alpha,\alpha'}\gg |t_\alpha|$) and trace the  evolution of the phase diagram with the increase in the degree of atomic nonidentity starting from nearly identical atoms. We show that nature of the ground states and the set of quantum phase transitions of sufficiently distinct atoms are qualitatively different from those in the case of (nearly) identical atoms considered earlier.

\begin{figure*}[t]
  \centering
  \includegraphics[width=\textwidth]{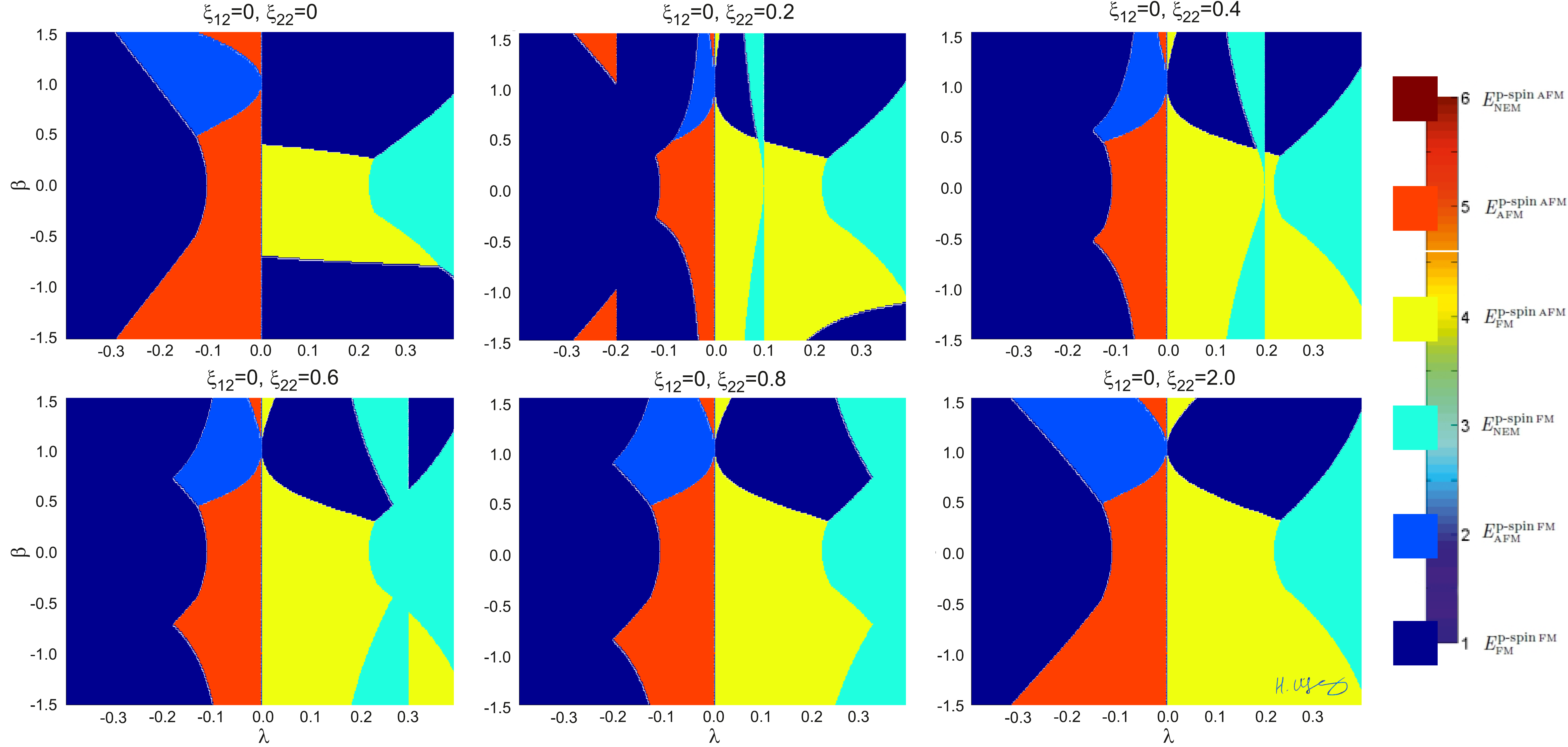}
  \caption{(Color online) Phase diagrams for $\xi_{12}=U_{12}/U_{11}=0$. Here $\beta= t_2/t_1$, $\xi_{22}=U_{22}/U_{11}$  and $\lambda=U_s/U_{11}$. In this case there is no Coulomb interaction between different atom kinds. The colorbar defines different spin and pseudospin orders.}\label{Fig_xi12_00_xi22_00-20}
\end{figure*}

Actually, different ultra cold atoms in optical lattices have been  considered in a number of papers in the last decade, see, e.g., Refs.~\onlinecite{Kuklov2003PhysRevLett,paredes2004tonks,Isacsson2005PhysRevB,
Das2007PhysRevA,Hofstetter2009PhysRevB,Takayoshi2010PhysRevA,Chen2012PhysRevB,
Chtchelkatchev2014PRA,Abanin2015PhysRevLett,belemuk2017PRB,Potter2017PhysRevLett,
Richaud2017PhysRevA,hu2018periodically}, and Refs.~\onlinecite{George14_RoMP,Lewenstein2015RepProgPhys,Gross2017science} for review. Numerous striking effects induced by the multispecie nature of boson system have been found including quantum phase transitions, many-body localization, and topological order,  as well as the superfluidity and supersolidity of ultracold atomic systems. However, most model investigations effectively deal with zero-spin boson species. Here, we focus on still unexplored physical phenomena in the systems with different species of vector bosons originating from the tunable interplay of spin degrees of freedom and of those identifying different sorts of atoms, see Fig.~\ref{Fig_xi12_00_xi22_00-20} for an illustration.

The rest of our paper is organized as follows: In Sec.~\ref{secModel}, we introduce the model Hamiltonian, then in Sec.~\ref{secHeff}, we reduce the initial general Hamiltonian for vector bosons to the effective Hamiltonian appearing to be anisotropic spin-pseudospin model of the Kugel--Khomskii type~\cite{Kugel1982UFN}; in Sec.~\ref{secEGround}, we investigate different possible configurations in spin and pseudospin spaces and find their energy; in Sec.~\ref{SecPhaseDiagram}, we discuss the energy of the ground state and the quantum phase transitions.  In particular, in Sec.~\ref{SecPhaseDiagram}, we illustrate the evolution of phase diagrams with the degree of atomic nonidentity, which looks like the transformation from the Joan Mir\'o style  artistic image to that of Kazimir Malewicz. Finally in Appendix, we present the analysis of several special limiting cases of the model Hamiltonian that relate our model system to some well-known results.

\section{Hamiltonian for two species of vector bosons} \label{secModel}

We consider two types of boson atoms with $S= 1$ in the optical lattice with sites labeled by index $i$. The corresponding creation operators $ c^{\dagger}_{i \alpha s}$ where $s= \{ -1, 0, 1 \}$ is the spin index and $\alpha= 1, 2$ labels the type of the boson.

The Hamiltonian includes three terms:
\begin{gather}
 H=H^{\scriptscriptstyle (U_0)}+ H^{\scriptscriptstyle (U_s)}+ H_t.
\end{gather}
The interaction between bosons is given by two terms:
\begin{gather}
H^{\scriptscriptstyle (U_0)}=\sum_iU_{12} n_{i,1}  n_{i,2} +\frac12\sum_{i,\alpha=1,2} U_{\alpha\alpha}  n_{i,\alpha} ( n_{i,\alpha}- 1).
\end{gather}
The first term corresponds to the repulsion between boson atoms at the same site~\cite{Lewenstein2015RepProgPhys}. Here, $U_{11}$, $U_{22}$, and $U_{12}$ are three interaction parameters and $ n_{i\alpha}=\sum_s c^{\dagger}_{i \alpha s} c_{i \alpha s}$.  Now we assume that the interaction constants can strongly differ from each other, contrary to the case considered in Ref.~\cite{belemuk2017PRB}.

The spin-dependent interaction term is taken in the standard form~\cite{Imambekov03,Tsuchiya04}:
\begin{gather}
 H^{\scriptscriptstyle (U_s)}_i= U_s( \mathbf S_i^2- 2n_i)/2,
\end{gather}
where $n_i= n_{i,1}+ n_{i,2}$ is the total number of bosons at site $i$.

The hopping term
\begin{gather} \label{hops}
  H_t= \sum_{\langle i,j \rangle}h_t \equiv \sum_{\langle i,j \rangle}t_{\alpha} \left( c^{\dagger}_{i \alpha \sigma} c_{j \alpha \sigma}+ c^{\dagger}_{j \alpha \sigma} c_{i \alpha \sigma} \right),
\end{gather}
where $\langle i,j \rangle$ means the summation only over the nearest-neighbor sites and $h_t$ is the hopping Hamiltonian for one link. As usual, the repeated indices imply summation.

In \eqref{hops} and further on the spin variable is $\sigma$, whereas  before the spin variable is defined as $s$.

\section{Effective Hamiltonian \label{secHeff}}
Below we focus on the Mott insulating state, where cold atoms are localized at the sites of optical lattice with the number of bosons at each site equal to unity. We remind that in such case, the hopping terms~\eqref{hops} can be treated as perturbation compared to the interaction part of the Hamiltonian, $H^{\scriptscriptstyle (U_0)}+ H^{\scriptscriptstyle (U_s)}$. Application of the perturbation theory reduces the initial general Hamiltonian to a simpler effective Hamiltonian written solely in terms of the spin and  pseudospin (p-spin) operators related to the lattice sites with atom filling equal to one. These spin-1 $\mathbf S_i$ and p-spin-1/2 $\bm{\mathcal T}_i$ operators are defined in a standard way~\cite{Yamashita1998PRB}
\begin{gather}\label{eqSc}
  S^a_i= c^\dag_{i\alpha \sigma} s^a_{\sigma\sigma'}c_{i\alpha \sigma'}, \qquad
  {\mathcal T}^a_i= c^\dag_{i\alpha \sigma} \tau^a_{\alpha\beta} c_{i\beta \sigma},
\end{gather}
where $a=x,y,z$. 

Below we  outline the algorithm of transforming the initial Hamiltonian to the effective one.

\subsection{Basis states}

In what follows, when we consider the link $\langle i, j \rangle$ between the nearest-neighbor sites, we focus on the basis of possible states for two bosons with spins $S_1= 1$ and $S_2= 1$ at neighboring sites $i=1$ and $j=2$. We are interested in the case with single occupation, i.e. when one boson of either type is located at each lattice site, $n_{i1}+ n_{i2}= 1$. We can pass now to the basis of the eigenstates of the total spin squared ${\bf S}^2= ({\bf S}_1+ {\bf S}_2)^2$ and its $z$-projection $S^z= S^z_1+ S^z_2$. We designate these states as $|S M \rangle$, $S= 0, 1, 2$ and $M=-S,\ldots,S$. This basis can be written as follows
\begin{gather}
\Phi^{(f)}_{SM}=  |\phi^{(f)}_S \rangle |S M \rangle, \label{basis1}
\end{gather}
where $f=1,\ldots, 4$ enumerates the ways to distribute two types of bosons over two sites. The coordinate part $ |\phi^{(f)}_S\rangle$ is given explicitly in Ref.~\cite{belemuk2017PRB}.

Applying $h_t$, see Eq.~\eqref{hops}, to the basis states \eqref{basis1}, we obtain two kinds of intermediate (virtual) states. The first type will be realized for two \textit{identical} bosons at one site ($i$ or $j$), the second type is for two \textit{nonidentical} bosons at one site.
Intermediate energies depend on the spin and types of bosons. They are
\begin{gather}
E^{aa}_{S= 0}= U_{11}- 2U_s, \quad E^{aa}_{S= 2}= U_{11}+ U_s, \notag
\\
E^{bb}_{S= 0}= U_{22}- 2U_s, \quad E^{bb}_{S= 2}= U_{22}+ U_s,
\\
E^{ab}_{S= 0}= U_{12}- 2U_s, \quad E^{ab}_{S= 2}= U_{12}+ U_s, \notag
\\
E^{ab}_{S= 1}= U_{12}- U_s.\notag
\end{gather}
There are no intermediate states corresponding to the total spin $S = 1$ for two identical $a$- or $b$-bosons due to the symmetry of the total wave function.

The energy of the virtual states with the double site occupancy is much larger than the energy of the states with the single site occupancy we are focusing at. To find corrections to the energy of the single-occupancy states related to the hoppings, we need the second order terms of the  perturbation theory. So, further it will be convenient to work with the operator
\begin{equation}
h= -h_{\rm eff}= h_t(1/H_0)h_t,
\end{equation}
where $H_0=H^{\scriptscriptstyle (U_0)}+ H^{\scriptscriptstyle (U_s)}$. In the basis of states \eqref{basis1}, the matrix of $h$ can be presented in the following block form
\begin{equation} \label{matrix_h}
h=
\begin{pmatrix}
B_{11} & 0      & 0      & 0 \\
0      & B_{22} & B_{23} & 0 \\
0      & B_{32} & B_{33} & 0 \\
0      & 0      & 0      & B_{44}
\end{pmatrix}.
\end{equation}
The matrix $h$ here is in fact the block-matrix, where each block is $9 \times 9$ matrix. Blocks $B_{kl}= \langle \Phi^k_{S'M'}|h|\Phi^l_{SM} \rangle$, $k, l= 1, 2, 3, 4$  are diagonal matrices. Their explicit forms are the following
\begin{equation}
B_{11}= 4t_1^2
\begin{pmatrix}
\frac{1}{E^{aa}_{S= 0}} I_1&  \\
                           & 0 \cdot I_3 &  \\
                           &            & \frac{1}{E^{aa}_{S= 2}} I_5 &
\end{pmatrix},
\end{equation}

\begin{equation}
B_{44}= 4t_2^2
\begin{pmatrix}
\frac{1}{E^{bb}_{S= 0}} I_1&  \\
                           & 0  \cdot I_3 &  \\
                           &            & \frac{1}{E^{bb}_{S= 2}} I_5 &
\end{pmatrix},
\end{equation}

\begin{equation}
B_{22}= B_{33}= (t_1^2+ t_2^2)
\begin{pmatrix}
\frac{1}{E^{ab}_{S= 0}} I_1&  \\
                           & \frac{1}{E^{ab}_{S= 1}} I_3 &  \\
                           &            & \frac{1}{E^{ab}_{S= 2}} I_5 &
\end{pmatrix},
\end{equation}

\begin{equation}
B_{23}= B_{32}= 2t_1 t_2
\begin{pmatrix}
\frac{1}{E^{ab}_{S= 0}} I_1&  \\
                           & \frac{1}{E^{ab}_{S= 1}} I_3 &  \\
                           &            & \frac{1}{E^{ab}_{S= 2}} I_5 &
\end{pmatrix}.
\end{equation}
Here $I_n$, $n=1,3,5$, are the identity $n \times n$ matrices. $I_1$ accounts for one state with $S= 0$, $I_3$ accounts for three states with $S= 1$, and $I_5$ accounts for five states with $S= 2$.

In what follows, we identify the single occupancy of site $i$  with $a$- or $b$-boson by the pseudospin-$ 1/2$ states $|1 \rangle_i= |+ \rangle_i$ and $|2 \rangle_i= |- \rangle_i$, respectively.  It will be convenient to introduce explicitly two types of creation operators: $a^{\dagger}_{is}$ for bosons of type $\alpha= 1$ and $b^{\dagger}_{is}$ for bosons of type $\alpha =2$. 

For this purpose, we rewrite the pseudospin operator ${\mathcal T}^\gamma_i$ at sites $i$, see Eq.~\eqref{eqSc}, in the form
\begin{equation}
{\mathcal T}^{\gamma}_i= a^{\dagger}_{i s} \tau^{\gamma}_{11} a_{i s}+ a^{\dagger}_{i s} \tau^{\gamma}_{12} b_{i s}+ b^{\dagger}_{i s} \tau^{\gamma}_{21} a_{i s}+ b^{\dagger}_{i s} \tau^{\gamma}_{22} b_{i s}.
\end{equation}
We can rewrite, as usual, the set of $\mathcal T^{\gamma}$ operators in the other equivalent form:
\begin{equation}\notag
T^{+}_i= a^{\dagger}_{i s} b_{i s}, \quad T^{-}_i= b^{\dagger}_{i s} a_{i s}, \quad T^{z}_i= \frac12 (a^{\dagger}_{i s} a_{i s}- b^{\dagger}_{i s} b_{i s}).
\end{equation}

To describe the occupancy of sites $i$ and $j$, we introduce the basis of pseudospin states $|\alpha \beta \rangle= |\alpha \rangle_i |\beta \rangle_j$.
Then, we find the correspondence between two-boson orbital states~\eqref{basis1} and pseudospin states $|\alpha \beta \rangle$. For example, $|\phi^{(1)}_S \rangle \longrightarrow |+ + \rangle$.

In what follows, we map the matrix $h$, Eq. \eqref{matrix_h}, onto an effective spin-pseudospin operator in the space $|\alpha \rangle_i |\beta \rangle_j |SM \rangle$. This operator will be given in terms of spin $S= 1$ operators ${\bf S}_i$, ${\bf S}_j$ and pseudospin $T= 1/2$ operators ${\bf T}_i$, ${\bf T}_j$ and it has the same structure as matrix $h$, Eq. \eqref{matrix_h}.

Next, we introduce the projection operator $Q_S$ and $P_T$ onto the combination of states $|S M \rangle$ and  $|T M_T\rangle$ corresponding to the total spin $S= 0, 1, 2$ and pseudospin $T= 0, 1$ at the link $\langle i,j \rangle$. The projectors in the spin space $Q_S= \sum_{M= -S}^{S} |S M \rangle \langle S M |$ can be written as
\begin{gather}  \notag
Q_0= -\frac{1}{3}+ \frac{1}{3} ({\bf S}_i \cdot {\bf S}_j)^2,
\\\label{eqQ}
Q_1= 1 -\frac{1}{2}({\bf S}_i \cdot {\bf S}_j)- \frac{1}{2}({\bf S}_i \cdot {\bf S}_j)^2,
\\ \notag
Q_2= \frac{1}{3}+ \frac{1}{2}({\bf S}_i \cdot {\bf S}_j)+ \frac{1}{6}({\bf S}_i \cdot {\bf S}_j)^2,
\end{gather}
where $Q_0+ Q_1+ Q_2= 1$.

Similarly, in the pseudospin space the projectors onto the singlet $T= 0$ and triplet $T= 1$ states are
\begin{gather}
P_{s}= \frac{1}{4}- {\bf T}_i \cdot {\bf T}_j, \quad
P_{t}= \frac{3}{4}+ {\bf T}_i \cdot {\bf T}_j .
\end{gather}

It is also convenient to introduce the following projectors in the pseudospin space

\begin{align}  \notag
P^{11}&= |+ + \rangle \langle + + |= \left(\frac{1}{2}+ T^z_i \right) \left(\frac{1}{2}+ T^z_j \right),
 \\ \notag
P^{22}&= |+ - \rangle \langle + - |= \left(\frac{1}{2}+ T^z_i \right) \left(\frac{1}{2}- T^z_j \right),
\\\label{eqP}
P^{33}&= |- + \rangle \langle - + |= \left(\frac{1}{2}- T^z_i \right) \left(\frac{1}{2}+ T^z_j \right),
\\ \notag
P^{44}&= |- - \rangle \langle - - |= \left(\frac{1}{2}- T^z_i \right) \left(\frac{1}{2}- T^z_j \right),
\end{align}
and
\begin{gather}\label{eqP32}
P^{32}= |- + \rangle \langle + - |= T^{-}_i T^{+}_j,
\\ \notag
P^{23}= |+ - \rangle \langle - + |= T^{+}_i T^{-}_j.
\end{gather}

In addition, we use below the following identities:
\begin{align} \notag
P^{11}+ P^{44}&= \frac{1}{2}+ 2 T^z_i T^z_j,
\\\label{eqPi}
P^{22}+ P^{33}&= \frac{1}{2}- 2 T^z_i T^z_j,
\\ \notag
P^{32}+ P^{23}&= 2 {\bf T}_i \cdot {\bf T}_j- 2 T^z_i T^z_j.
\end{align}

\begin{figure*}[t]
  \centering
  \includegraphics[width=\textwidth]{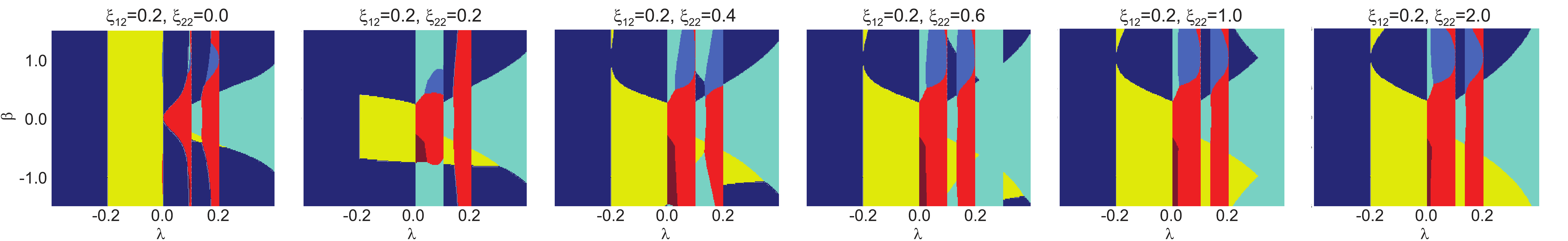}
  \caption{(Color online) Phase diagrams for small Coulomb interaction between different atom kinds: $\xi_{12}=U_{12}/U_{11}=0.2$. All the notations follow Fig.~\ref{Fig_xi12_00_xi22_00-20}. \label{Fig2_xi12_02_xi22_00-20}}
\end{figure*}
\begin{figure*}[t]
  \centering
  \includegraphics[width=\textwidth]{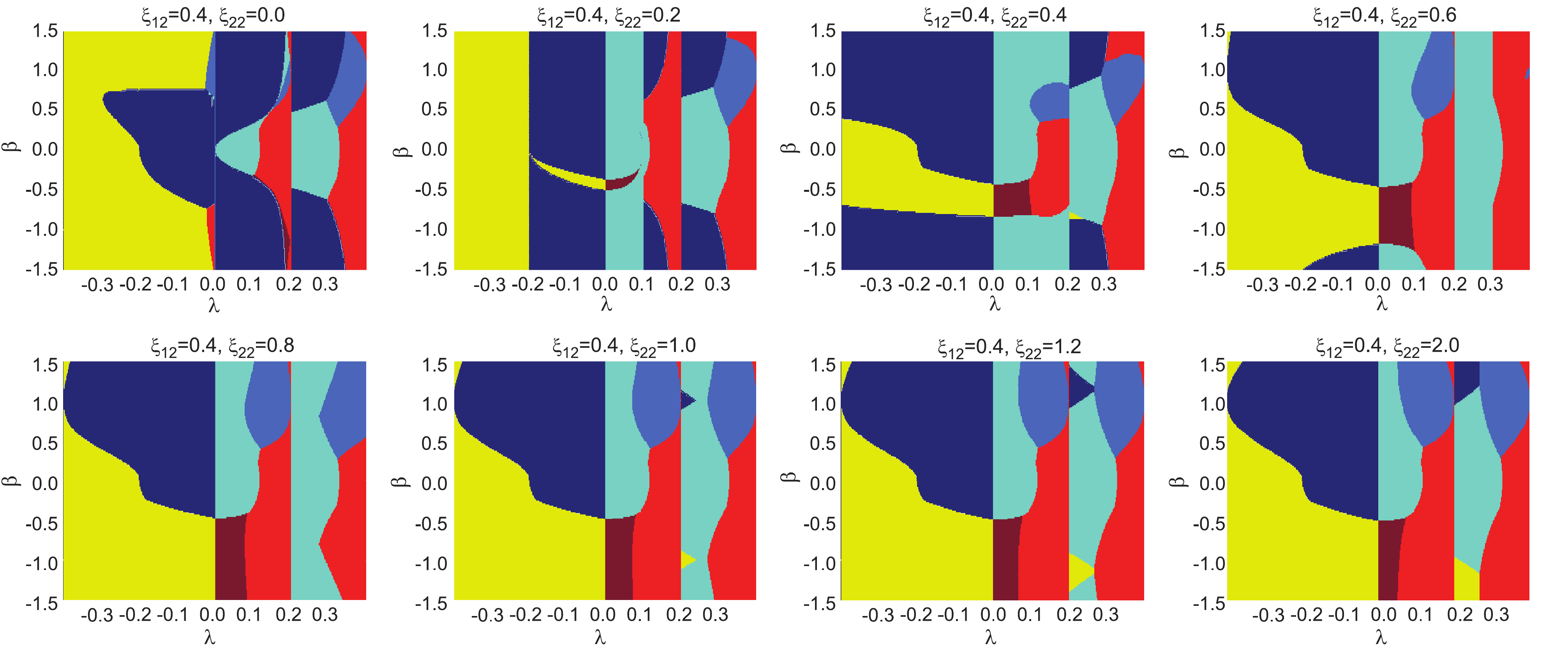}
  \caption{(Color online) Phase diagrams for $\xi_{12}=U_{12}/U_{11}=0.4$. 
  }
\end{figure*}

With the help of the projectors~\eqref{eqQ}--\eqref{eqP32} and the identifie~\eqref{eqPi}, we rewrite the block matrix $h$ in terms of spin and pseudospin operators as follows
\begin{widetext}\begin{multline}
h= 4t_1^2 {\,} P^{11} \left[ \frac{Q_0}{E^{aa}_0}+ \frac{ Q_2}{E^{aa}_2} \right]+ 4t_2^2 {\,} P^{44} \left[ \frac{Q_0}{E^{bb}_0}+ \frac{ Q_2}{E^{bb}_2} \right]+ (t_1^2+ t_2^2) \left[ P^{22}+ P^{33} \right] \left[ \frac{ Q_0}{E^{ab}_0}+ \frac{ Q_1}{E^{ab}_1}+  \frac{ Q_2}{E^{ab}_2} \right]+ \\
+ 2t_1 t_2 \left[ P^{23}+ P^{32} \right] \left[ \frac{Q_0}{E^{ab}_0}+ \frac{Q_1}{E^{ab}_1}+  \frac{ Q_2}{E^{ab}_2} \right].
\end{multline}
Substituting the explicit form of pseudospin projectors, we can rewrite $h$ in the form containing only spin projectors

\begin{multline} \label{h_gen1}
h= \left\{ \frac{t_1^2}{E^{aa}_0}+ \frac{t_2^2}{E^{bb}_0}+ 2 \left( \frac{t_1^2}{E^{aa}_0}- \frac{t_2^2}{E^{bb}_0}\right) (T^z_i+ T^z_j)+ 4 \left( \frac{t_1^2}{E^{aa}_0}+ \frac{t_2^2}{E^{bb}_0}\right) T^z_i T^z_j \right\} Q_0+ \\
+ \left\{ \frac{t_1^2}{E^{aa}_2}+ \frac{t_2^2}{E^{bb}_2}+ 2 \left( \frac{t_1^2}{E^{aa}_2}- \frac{t_2^2}{E^{bb}_2}\right) (T^z_i+ T^z_j)+ 4 \left( \frac{t_1^2}{E^{aa}_2}+ \frac{t_2^2}{E^{bb}_2}\right) T^z_i T^z_j \right\} Q_2+ \\
+ \left\{ \frac{1}{2}(t_1^2+ t_2^2)- 2(t_1+ t_2)^2 T^z_i T^z_j+ 4t_1 t_2 {\,} {\bf T}_i \cdot {\bf T}_j \right\}  \left[ \frac{Q_0}{E^{ab}_0} + \frac{Q_1}{E^{ab}_1} +  \frac{Q_2}{E^{ab}_2} \right].
\end{multline}
\end{widetext}

Finally, the effective Hamiltonian is written as the sum of $h$ operators \eqref{h_gen1} over all $\langle i,j \rangle$ links
\begin{equation}\label{heff}
H_{\rm eff}=- \sum \limits_{\langle i,j \rangle} h .
\end{equation}
This is the most general form of the effective Hamiltonian involving  different Hubbard interaction parameters, different hopping amplitudes, and spin-dependent interaction. In Appendix, we show that in a number of  limiting cases, this Hamiltonian can be simplified to some well-known forms.

\begin{figure*}[t]
  \centering
  \includegraphics[width=\textwidth]{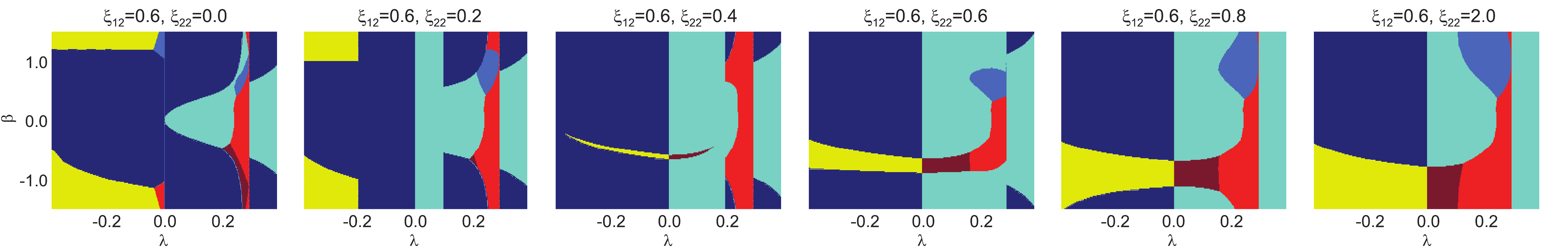}
  \caption{(Color online) Phase diagrams for $\xi_{12}=U_{12}/U_{11}=0.6$.}
\end{figure*}

\begin{figure*}[t]
  \centering
  \includegraphics[width=\textwidth]{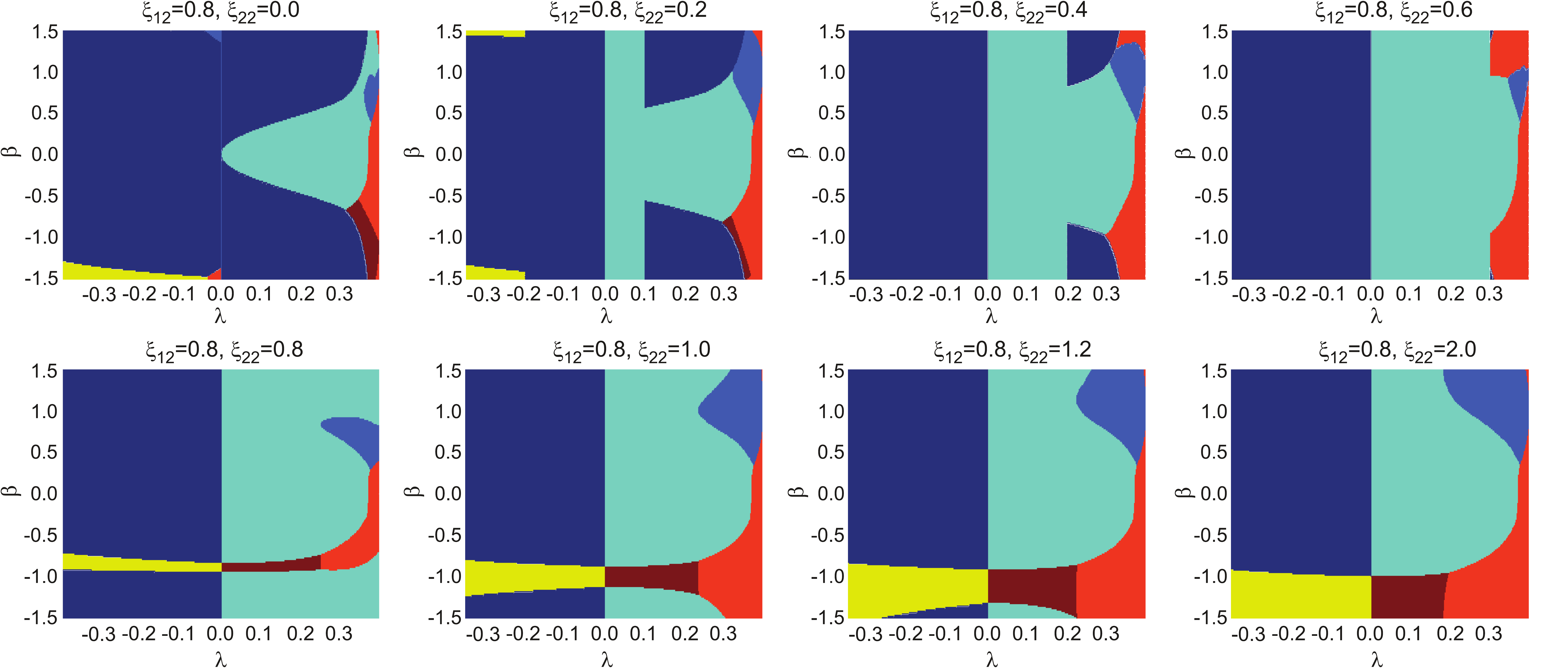}
  \caption{(Color online) Phase diagrams for $\xi_{12}=U_{12}/U_{11}=0.8$.}
\end{figure*}

\section{Energy of the ground state~\label{secEGround}}

To proceed with the calculation of the ground state energy, we rewrite the ``kernel'' $h$ of effective Hamiltonian~\eqref{heff} as follows
\begin{gather}
h= \frac{2t_1^2}{U_{11}}\left( R_0 Q_0+ R_2 Q_2+ R_1 \left[ \frac{Q_0}{E_0^{ab}}+ \frac{Q_1}{E_1^{ab}}+
\frac{Q_2}{E_2^{ab}}\right] \right),
\end{gather}
where the coefficients $R_0$, $R_1$ and $R_2$ are
\begin{multline}
R_0= \left[ \frac12 \left(\frac{1}{E_0^{aa}}+ \frac{\beta^2}{E_0^{bb}}\right)+ \left(\frac{1}{E_0^{aa}}-
\frac{\beta^2}{E_0^{bb}}\right) (T^z_i+ T^z_j)+\right.
\\ \left.
 2 \left(\frac{1}{E_0^{aa}}+ \frac{\beta^2}{E_0^{bb}}\right)
(T^z_i T^z_j) \right],
\end{multline}
\begin{equation}
R_1= \left[ \frac14(1+ \beta^2)- (1+ \beta)^2 (T^z_i T^z_j)+ 2\beta ({\bf T}_i \cdot {\bf T}_j) \right],
\end{equation}
\begin{multline}
R_2= \left[ \frac12 \left(\frac{1}{E_2^{aa}}+ \frac{\beta^2}{E_2^{bb}}\right)+ \left(\frac{1}{E_2^{aa}}-
\frac{\beta^2}{E_2^{bb}}\right) (T^z_i+ T^z_j)+\right.
\\ \left.
2 \left(\frac{1}{E_2^{aa}}+\frac{\beta^2}{E_2^{bb}}\right) (T^z_i T^z_j) \right].
\end{multline}
Here, we introduce the dimensionless parameter $\beta=t_2/t_1$ that characterizes the difference of the tunnel amplitudes for different boson species.
\begin{figure*}[t]
  \centering
  \includegraphics[width=\textwidth]{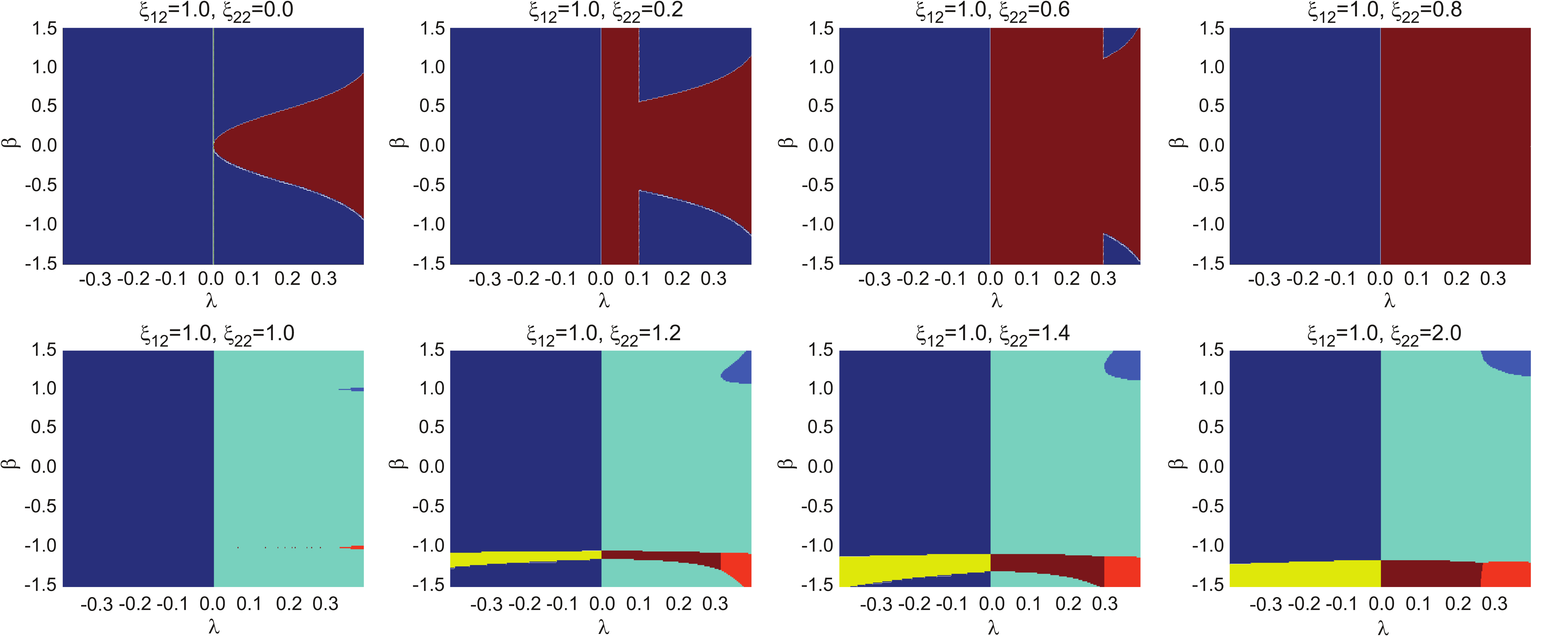}
  \caption{(Color online) Phase diagrams for $\xi_{12}=U_{12}/U_{11}=1.0$.}
\end{figure*}
\begin{figure*}[t]
  \centering
  \includegraphics[width=\textwidth]{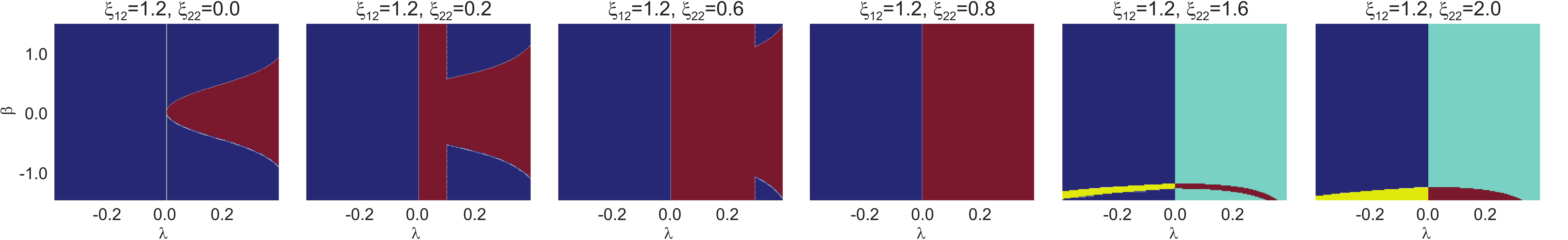}
  \caption{(Color online) Phase diagrams for $\xi_{12}=U_{12}/U_{11}=1.2$.}
\end{figure*}

The energy $E$ of the ground state can be formally written as the average of the effective Hamiltonian over the ground state wave function (we will find it later using the variational approach)
\begin{gather}\label{eqEmf}
E=- \frac{\nu}{2} \langle h \rangle,
\end{gather}
where  $\nu= 2D$ is the number of nearest neighbors for the $D$-dimensional cubic lattice.

Within the mean-field approximation, we can neglect any correlations between spin and p-spin degrees of freedom. Then, for example, $ \langle R_0 Q_0 \rangle\to  \langle R_0\rangle \langle Q_0 \rangle$, and
\begin{multline}
\langle h \rangle= E_u\left(  \langle R_0\rangle \langle Q_0 \rangle +\langle R_2\rangle \langle Q_2 \rangle+ \right.
\\\left.
\langle R_1 \rangle \left[ \frac{\langle Q_0 \rangle }{E_0^{ab}}+ \frac{\langle Q_1 \rangle}{E_1^{ab}}+
\frac{\langle Q_2 \rangle}{E_2^{ab}}\right] \right),
\end{multline}
where $ E_u= \frac{2t_1^2}{U_{11}}$.

We take the trial wave function in the p-spin space as the ``mixed orbital state'' on two-sublattices $A$ and $B$
\begin{gather}
|\chi \rangle_i= \cos \theta |+\rangle_i+ \sin \theta |-\rangle_i, \quad i \in A, \\
|\chi \rangle_j= \cos \theta |+\rangle_j+ \eta \sin \theta |-\rangle_j, \quad j \in B, \quad \eta= \pm 1.
\end{gather}
For $\theta= 0$, it defines the p-spin ferromagnetic (FM) state
\begin{equation}
|\uparrow_i \uparrow_j \uparrow_i \uparrow_j \dots \rangle.
\end{equation}
For $\theta= \frac{\pi}{2}$, it defines also the p-spin FM state
\begin{equation}
|\downarrow_i (-\downarrow_j) \downarrow_i (-\downarrow_j) \dots \rangle.
\end{equation}

The FM and antiferromagnetic (AFM) p-spin states will occur at  $\theta= \frac{\pi}{4}$ for $\eta= +1$ and $\eta= -1$, respectively. These states are polarized in $x$ direction:
\begin{gather}
\frac{1}{\sqrt{2}}(|\uparrow \rangle+  |\downarrow \rangle)_i, \\
\frac{1}{\sqrt{2}}(|\uparrow \rangle+ \eta |\downarrow \rangle)_j
\end{gather}

For other angles $\theta$, the mixed orbital state is a some intermediate state between the FM and AFM ones. That is why, it is more correctly to refer to this mixed orbital state as the mixed orbital state with $\eta= +1$, or $\eta= -1$ rather than the FM and AFM states.

\begin{figure*}[t]
  \centering
  \includegraphics[width=\textwidth]{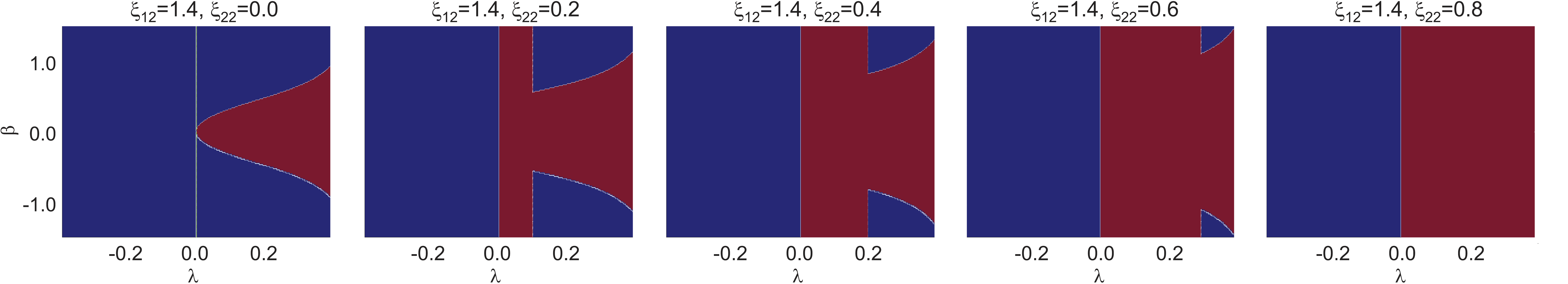}
  \caption{(Color online) Phase diagrams for $\xi_{12}=U_{12}/U_{11}=1.4$.}
\end{figure*}
\begin{figure*}[t]
  \centering
  \includegraphics[width=0.62\textwidth]{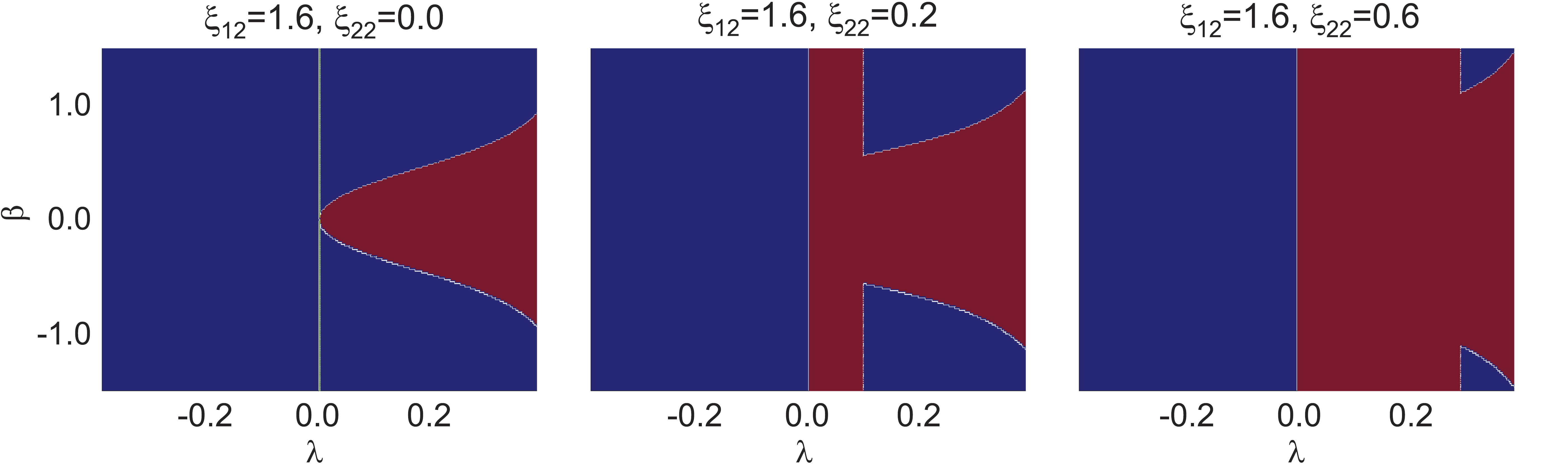}
  \caption{(Color online) Phase diagrams for $\xi_{12}=U_{12}/U_{11}=1.6$.\label{Fig9_xi12_16_xi22_00-20}}
\end{figure*}

Now, we find averages over the trial mixed orbital state of the p-spin operators entering $R_i$, $i=0,1,2$
\begin{equation}
\langle T^z_i+ T^z_j \rangle= \left\{
\begin{aligned}
&\cos 2\theta, \quad &\eta= +1, \\
&0, \quad &\eta= -1,
\end{aligned} \right.
\end{equation}
\begin{equation}
\langle T^z_i T^z_j \rangle= \left\{
\begin{aligned}
&\frac14 \cos^2 2\theta, \quad &\eta= +1, \\
&-\frac14 \cos^2 2\theta, \quad &\eta= -1,
\end{aligned} \right.
\end{equation}
\begin{equation}
\langle {\bf T}_i {\bf T}_j \rangle= \left\{
\begin{aligned}
&\frac14, \quad &\eta= +1, \\
&-\frac14, \quad &\eta= -1.
\end{aligned} \right.
\end{equation}
Later on, we will minimize the energy $E$ with respect to the angle $\theta$.

In contrast to the p-spin space, the effective Hamiltonian in the spin space is isotropic. So, we do not expect any exotic states there. Hence, we take for trial wave functions in the spin space the usual FM, AFM, and nematic (NEM) states~\cite{Imambekov03}.

Then, the coefficients for projectors in the spin space are
\begin{gather}\notag
\langle Q_0 \rangle= \left\{
\begin{aligned}
&\,0\,, \quad \mathrm{FM}, \\
&\frac13, \quad \mathrm{AFM}, \\
&\frac13, \quad \mathrm{NEM},
\end{aligned} \right. \qquad
\langle Q_1 \rangle= \left\{
\begin{aligned}
&\,0\,, \quad \mathrm{FM}, \\
&\frac12 , \quad \mathrm{AFM}, \\
&\,0\,, \quad \mathrm{NEM},
\end{aligned} \right.
\\
\langle Q_2 \rangle= \left\{
\begin{aligned}
&\,1\,, \quad \mathrm{FM}, \\
&\frac16, \quad \mathrm{AFM}, \\
&\frac23, \quad \mathrm{NEM}.
\end{aligned} \right.
\end{gather}

Finally, we have found all the averages and correlation functions entering Eq.~\eqref{eqEmf} for the energy. We minimize numerically the energy  $E(\theta,\eta)$, find the ground stats at different values of the parameters, and draw the corresponding phase diagrams.

\section{Results and discussion~\label{SecPhaseDiagram}}

\subsection{General properties of the phase diagrams}

We address here the ground state of the system and possible quantum phase transitions. Performing the energy minimization, we find the phase diagrams for different ranges of the parameters. The key parameters are $\beta= t_2/t_1$, $\lambda=U_s/U_{11}$, $\xi_{12}=U_{12}/U_{11}$, and $\xi_{22}=U_{22}/U_{11}$. All these parameters  (except $\lambda$ responsible for spin channel interaction) are related to the difference between the types of bosons.

There are six different phases: three phases have the ferromagnetic pseudospin arrangement, while three others correspond to the  antiferromagnetic pseudospin state. Evolution of these phases is illustrated in Figs.~\ref{Fig_xi12_00_xi22_00-20}--\ref{Fig9_xi12_16_xi22_00-20};   Fig.~\ref{Fig_xi12_00_xi22_00-20} is supplemented by the colorbar, where the correspondence between colors and phases is shown. Points at the colorbar correspond to the following orders in spin and pseudospin systems:
\begin{eqnarray}
\notag
1 &\rightarrow&\EFMpFM\,,
\\\notag
2 &\rightarrow&\EAFMpFM\,,
\\\label{eqsEnergies}
3 &\rightarrow&\ENEMpFM \,,
\\\notag
4 &\rightarrow&\EFMpAFM\,,
\\\notag
5 &\rightarrow&\EAFMpAFM\,,
\\\notag
6 &\rightarrow&\ENEMpAFM \,.
\end{eqnarray}
Here, for example, color ``1'' corresponds to ferromagnetic spin and pseudospin orders.

In Figs.~\ref{Fig_xi12_00_xi22_00-20}--\ref{Fig9_xi12_16_xi22_00-20}, we demonstrate the evolution of the phase diagrams within the wide range of parameters $\beta= t_2/t_1$, $\lambda=U_s/U_{11}$, $\xi_{12}=U_{12}/U_{11}$ and $\xi_{22}=U_{22}/U_{11}$. The most interesting phase transition is that accompanied by the change of atom distribution over the optical lattice: p-spin FM $\leftrightarrow$ p-spin AFM. Note that the cold colors correspond to FM p-spin, while the warm ones --- to p-spin AFM. So the transitions with p-spin change can be found in the phase diagram at the lines, where cold colors change to warm ones.

Evolution of the phase diagrams for $\xi_{12}=U_{12}/U_{11}=0$ with $\xi_{22}=U_{22}/U_{11}$ on $(\lambda,\beta)$-plane is shown in Fig.~\ref{Fig_xi12_00_xi22_00-20}. In this case, there is no Coulomb interaction between different  boson species. We see that there is always quantum phase transition at the line $\lambda=0$. This is true not only for Fig.~\ref{Fig_xi12_00_xi22_00-20}, but also for all phase diagrams in Figs.~\ref{Fig_xi12_00_xi22_00-20}-\ref{Fig9_xi12_16_xi22_00-20}. This phase transition is driven by the sign change of spin channel interaction $U_s$. This transition has been recently revealed in Ref.~\cite{belemuk2017PRB} for nearly identical vector bosons. Here, we show that this transition is very stable with respect to the evolution of the degree of nonidentity.

One can also see in Fig.~\ref{Fig_xi12_00_xi22_00-20}, that the ``left color'' is always blue. It corresponds to FM spin ordering (with FM p-spin ordering). This situation is intuitively obvious: since the ``left'' phase is determined by large negative $\lambda$ --- the interaction in the spin channel.  In all the next figures, the ``left color'' also corresponds to the FM spin ordering, sometimes with the AFM p-spin ordering (yellow).

Looking through the complete set of phase  diagrams, one can notice the absence of exact symmetry with the respect to reflections $\lambda \to -\lambda$ and $\beta \to -\beta$, though some traces are detectable. The $\beta$-symmetry is restored, when $\xi_{12} \gg \xi_{22}$.

Let us mention that the intuitive speculations useful, for example, in the case of simple Heisenberg model, can be misleading here, because the effective Hamiltonian is rather nontrivial.

We also underline the evolution of the artistic image of the phase diagrams. Namely, at small $\xi_{12}$ their style resembles the J.~Mir\'o paintings, while at large $\xi_{12}$ ---  those of K.~Malewicz.

\subsection{Phase diagrams: specific features}

Figures~\ref{Fig_xi12_00_xi22_00-20}--\ref{Fig2_xi12_02_xi22_00-20} are the most multicolored --- there are quantum phase transitions nearly between all the possible phases. These pictures correspond to the low or moderate interspecies Coulomb interaction $U_{12}$ as compared to the single-species one. This feature can be attributed to small or moderate difference in the parameters characterizing their nonidentity. One can also notice that there are many reentrant phase transitions in Figs.~\ref{Fig_xi12_00_xi22_00-20}--\ref{Fig9_xi12_16_xi22_00-20}, especially for moderate $\xi_{12}$.

When $\xi_{22}$ becomes sufficiently large, then the nematic (NEM) spin phase prevails. The $\xi_{22}$-threshold for this behavior is the smallest at large $\xi_{12}$, as can be seen in Figs.~\ref{Fig_xi12_00_xi22_00-20}--\ref{Fig9_xi12_16_xi22_00-20}.

\section{Conclusions}

To conclude, we have investigated the evolution of the quantum state of vector two-species bosons in optical lattices with the  ``degree of atomic nonidentity'' that drives the cascade of quantum phase transitions. We have transferred the initial general Hamiltonian for vector bosons to the  anisotropic spin-pseudospin model of the Kugel--Khomskii type that served as the effective Hamiltonian. The variational approach have been used to uncover the phase diagram of the system in hand. We have investigated also limiting cases of the effective Hamiltonian and demonstrated the relation of our rather complicated Hamiltonian to the well known results.

\begin{acknowledgments}
A.M. and A.B. are grateful to Wu-Ming Liu for interest to this work and hospitality in Beijing National Laboratory for Condensed Matter Physics. N.C. is grateful to A.~Pekovic for stimulating discussions at the initial stage of this work, to Laboratoire de Physique Th\'eorique, Toulouse, where this work has been initiated, for the hospitality, and to CNRS.

We express our gratitude to the Computational Centers of Russian Academy of Sciences and National Research Center Kurchatov Institute for providing the access to URAL, JSCC, and HPC facilities.  This work was supported by the Russian Foundation for Basic Research (projects Nos. 16-02-00295, 16-02-00304, 17-02-00135, 17-52-53014, and 17-02-00323). The research of N.C. was made possible by Government of the Russian Federation (Agreement No. 05.Y09.21.0018).
\end{acknowledgments}

\appendix

\section{Special cases of Hamiltonian \eqref{h_gen1}--\eqref{heff}}
We remind that $H_{\rm eff}= -\sum \limits_{\langle i,j \rangle} h$.

\subsection{Equal Hubbard interaction parameters}

Now, we take a look at more special cases. First we consider the case with equal Hubbard interaction parameters
\begin{equation}
U_{12}= U_{11}= U_{22}= U_0
\end{equation}
Then
\begin{gather}
E^{ab}_S= E^{aa}_S= E^{bb}_S= E_S, \qquad S=0,1,2,
\end{gather}
and as follows from Eq.~\eqref{h_gen1}, the effective matrix $h$ reduces to the following form:
\begin{widetext}
\begin{multline} \label{h_gen2}
h= \frac{Q_0}{E_0} \left\{ \frac{3}{2} (t_1^2+ t_2^2)+ 2(t_1^2- t_2^2) (T^z_i+ T^z_j)+ 2(t_1- t_2)^2 {\,} T^z_i T^z_j+ 4 t_1 t_2 {\,} {\bf T}_i \cdot {\bf T}_j \right\} + \\
+ \frac{Q_2}{E_2} \left\{ \frac{3}{2} (t_1^2+ t_2^2)+ 2(t_1^2- t_2^2) (T^z_i+ T^z_j)+ 2(t_1- t_2)^2 {\,} T^z_i T^z_j+ 4 t_1 t_2 {\,} {\bf T}_i \cdot {\bf T}_j  \right\} + \\
+ \frac{Q_1}{E_1} \left\{ \frac{1}{2} (t_1^2+ t_2^2)- 2(t_1+ t_2)^2 {\,} T^z_i T^z_j+ 4 t_1 t_2 {\,} {\bf T}_i \cdot {\bf T}_j  \right\}
\end{multline}
\end{widetext}

\subsection{Equal Hubbard interaction parameters and equal hopping amplitudes}
\begin{widetext}
The effective matrix $h$ can be simplified further if one considers equal hopping amplitudes, $t_1= t_2= t$. For this case, as it can be seen from Eq.~\eqref{h_gen2}, matrix $h$ reduces to
\begin{equation}
h= \frac{Q_0}{E_0} \left\{ 3t^2+ 4 t^2 {\,} {\bf T}_i \cdot {\bf T}_j \right\} + \frac{Q_2}{E_2} \left\{ 3t^2+ 4 t^2 {\,} {\bf T}_i \cdot {\bf T}_j \right\}+ \frac{Q_1}{E_1} \left\{ t^2- 8 t^2 {\,} T^z_i T^z_j+ 4 t^2 {\,} {\bf T}_i \cdot {\bf T}_j \right\}.
\end{equation}
\end{widetext}

The last expression can be simplified if we take into account that
\begin{equation}
t^2- 8 t^2 {\,} T^z_i T^z_j+ 4 t^2 {\,} {\bf T}_i \cdot {\bf T}_j= 4t^2 {\,} P_{t0},
\end{equation}
where $P_{t0}= P_t- (P^{11}+ P^{44})= \left[ 1/4- 2 T^z_i T^z_j+ {\bf T}_i \cdot {\bf T}_j \right] $ is the projector onto the pseudospin state $|T= 1, M_T= 0 \rangle$, when the the matrix $h$ can be reduced to the form
\begin{equation} \label{h_3}
h= 4t^2 {\,} \left\{ \frac{Q_0}{E_0} P_t+ \frac{Q_2}{E_2} P_t+ \frac{Q_1}{E_1} P_{t0} \right\}.
\end{equation}

Note that due to the presence of the projector $P_{t0}$, the states with spin $S= 1$ will be automatically symmetric in the orbital space,
while the antisymmetric combination of orbital states is automatically excluded from the effective Hamiltonian.

Next, we rewrite the Hamiltonian $h$  in terms of the spin operators. We use the relation
\begin{widetext}
\begin{multline}
-4t^2 {\,} \left( \frac{Q_0}{E_0}+ \frac{Q_2}{E_2} \right)= \left\{ \frac{-4t^2}{3} \left( \frac{1}{E_2}- \frac{1}{E_0} \right)+  \frac{-4t^2}{2E_2} {\,} {\bf S}_i \cdot {\bf S}_j+ \frac{-4t^2}{3} \left( \frac{1}{E_0}+ \frac{1}{2E_2} \right){\,} ({\bf S}_i \cdot {\bf S}_j)^2 \right\}= \\
= \epsilon +  J {\,} {\bf S}_i \cdot {\bf S}_j+ K {\,} ({\bf S}_i \cdot {\bf S}_j)^2,
\end{multline}
\end{widetext}
where
\begin{gather}
J= \frac{-2t^2}{E_2}, \quad K= \frac{-4t^2}{3} \left( \frac{1}{E_0}+ \frac{1}{2E_2} \right),
\\
\epsilon= \frac{-4t^2}{3} \left( \frac{1}{E_2}- \frac{1}{E_0} \right)= J - K.
\end{gather}

Parameters $J$, $K$, and $\epsilon$ can be expressed explicitly via the initial interaction constants $U_0$ and $U_s$ as follows
\begin{gather}
J= \frac{-2t^2}{U_0+ U_s}, \quad K= \frac{-2t^2 U_0}{(U_0+ U_s)(U_0- 2U_s)},
\\
 \epsilon= \frac{4t^2 U_s}{(U_0+ U_s)(U_0- 2U_s)},
\end{gather}

\begin{widetext}
Now, the Hamiltonian takes the form
\begin{equation} \label{H_eff2}
H_{\rm eff}= \sum \limits_{\langle i,j \rangle} \left\{  \left[ \epsilon +  J {\,} {\bf S}_i \cdot {\bf S}_j+ K {\,} ({\bf S}_i \cdot {\bf S}_j)^2 \right] P_t+ \left( \frac{-4t^2}{E_1} \right) Q_1 P_{t0} \right\}.
\end{equation}

Finally, the relation $P_t= P_{t0}+ P^{11}+ P^{44}$ and the definition of $Q_1$~\eqref{eqQ}  allows us to reduce the Hamiltonian to the form used in Ref.~\cite{belemuk2017PRB},
\begin{equation} \label{H_eff3}
H_{\rm eff}= \sum \limits_{\langle i,j \rangle} \Bigl\{  \left[ \epsilon +  J {\,} {\bf S}_i \cdot {\bf S}_j+ K {\,} ({\bf S}_i \cdot {\bf S}_j)^2 \right] (P^{11}+ P^{44})+ \left[ \epsilon' +  J' {\,} {\bf S}_i \cdot {\bf S}_j+ K' {\,} ({\bf S}_i \cdot {\bf S}_j)^2 \right] {\,}P_{t0} \Bigr\},
\end{equation}
\end{widetext}
where we introduced
\begin{gather}
J'= 2t^2 \left(\frac{1}{E_1}- \frac{1}{E_2} \right),
\\
K'= \frac{-2t^2}{3} \left( \frac{2}{E_0}+ \frac{1}{E_2}- \frac{3}{E_1} \right),
\\
\epsilon'= \frac{-4t^2}{3} \left( \frac{1}{E_2}- \frac{1}{E_0}+ \frac{3}{E_1} \right).
\end{gather}

\subsubsection{One type of bosons}
For the case of only one type of bosons, Hamiltonian \eqref{H_eff3} is equivalent to that of Ref.~\cite{Yip03}. For single type of bosons, the pseudospin operators become $c$-numbers: $T^z_i= \pm 1/2$, ${\bf T}_i \cdot {\bf T}_j$ should be replaced by $T^z_i T^z_j= 1/4$, and projectors $P^{11}+ P^{44}= 1$, $P_{t0}= 0$. Then Eq.~\eqref{H_eff3} is reduced to the Hamiltonian considered in Ref.~\cite{Yip03}
\begin{equation}
H_{\rm eff}= \sum \limits_{\langle i,j \rangle} \left\{   \epsilon +  J {\,} {\bf S}_i \cdot {\bf S}_j+ K {\,} ({\bf S}_i \cdot {\bf S}_j)^2 \right\}.
\end{equation}

\subsubsection{No spin-dependent interaction}
If there is no spin-dependent interaction, i.e. $U_s= 0$, then $E_0= E_2= E_1$ and
$J= K= -2t^2/U_0$, $\epsilon= 0$, $J'= K'= 0$, and $\epsilon'= -4t^2/U_0= 2J$.
The effective Hamiltonian, Eq.~\eqref{H_eff3}, is reduced to
\begin{equation}
H_{\rm eff}= J  \sum \limits_{\langle i,j \rangle} \left\{ 2P_{t0}+ \bigl( {\bf S}_i \cdot {\bf S}_j+ ({\bf S}_i \cdot {\bf S}_j)^2 \bigr) \left(P^{11}+ P^{44}\right) \right\}.
\end{equation}

And at last, if in this case, there is only one type of bosons, then $P_{t0}= 0$, $P^{11}+ P^{44}= 1$, and the effective Hamiltonian describing effective interaction between identical bosons has the ferromagnetic character ($J < 0$)
\begin{equation}
H_{\rm eff}= J \sum \limits_{\langle i,j \rangle} \left\{ {\bf S}_i \cdot {\bf S}_j + ({\bf S}_i \cdot {\bf S}_j)^2 \right\}, \quad J= \frac{-2t^2}{U_0}.
\end{equation}
The  systems with identical bosons with odd and even number bosons per site were discussed in Refs.~\cite{Imambekov03,Tsuchiya04}.

\bibliography{mybibfile}
\end{document}